\documentclass[12pt,reqno]{article}
\usepackage{amsmath}
\usepackage{amsthm}
\usepackage{amsfonts}
\usepackage{amssymb}
\usepackage{graphicx,color}
\input{xy}
\xyoption{all}

\numberwithin{equation}{section}

\def\a{\alpha}
\def\b{\beta}
\def\c{\gamma}
\def\d{\delta}
\newcommand\e{\epsilon}
\newcommand\ad{{[1/2]}}

\def\sfrac12{{\scriptstyle \frac12}}

\def\P{{\mathcal{P}}}
\newcommand{\aP}{{\mathcal{\hat{P}}}}

\newcommand{\Complex}{\mathbb{C}}

\newcommand{\mat}{\begin{pmatrix}}
\newcommand{\tam}{\end{pmatrix}}

\newcommand{\mc}{\mathcal}

\newcommand\id{\text{id}}
\newcommand\tr{\text{tr}}
\newcommand\str{\text{str}}

\newcommand\WZNW{{\text{WZNW}}}
\newcommand\g{\mathfrak{g}}

\newcommand{\sll}{{\text{sl}(2)}}

\newcommand\PSL{{\text{PSL}(2|2)}}

\newcommand\psl{{\text{psl}(2|2)}}
\newcommand\psu{{\text{psu}(1,1|2)}}

\newcommand\s{{\sigma}}

\title{\bf Boundary spectra in superspace $\sigma$-models}
\author{\\[5mm] Thomas Quella$^{2,3}$,
               Volker Schomerus$^{1,3}$ and Thomas Creutzig$^{1}$\\[5mm]
$^1$ DESY Hamburg, Theory Group, \\
Notkestrasse 85, D--22607 Hamburg, Germany
\\[5mm]
$^2$ Institute for Theoretical Physics, University of Amsterdam,\\
Valckenierstraat 65, 1018 XE Amsterdam,
The Netherlands\\[5mm]
$^3$ Isaac Newton Institute for Mathematical Sciences,\\
20 Clarkson Road, Cambridge, CB3 0EH, United Kongdom\\[5mm]}
\begin{document}
\begin{titlepage}      \maketitle       \thispagestyle{empty}

\vskip1cm
\begin{abstract}
In this note we compute exact boundary spectra for D-instantons in
$\sigma$-models on the supergroup $\PSL$. Our results are obtained
through an explicit summation of the perturbative expansion for
conformal dimensions to all orders in the curvature radius. The
analysis exploits several remarkable properties of the
perturbation series that arises from rescalings of the metric on
$\PSL$ relative to a fixed Wess-Zumino term. According to
Berkovits, Vafa and Witten, the models are relevant in the context
of string theory on $AdS_3$ with non-vanishing RR-flux. The note
concludes with a number of comments on various possible generalizations
to other supergroups and higher dimensional supercoset theories.
\end{abstract}
\vspace*{-19.9cm} {\tt {DESY 07-226}} \hfill {\tt {yymm.nnnn}}\\
{\tt {ITFA-2007-55}} 
\bigskip\vfill
\noindent \phantom{wwwx}{\small e-mail: }{\small\tt
tquella@science.uva.nl, volker.schomerus@desy.de, thomas.creutzig@desy.de}
\end{titlepage}

\baselineskip=19pt \setcounter{equation}{0} \tableofcontents

\section{Introduction}

The celebrated AdS/CFT correspondence \cite{Maldacena:1998re,Aharony:1999ti}
has promoted the solution of string theory in Anti-de\,Sitter (AdS) spaces
to one of the central problems of modern mathematical physics. Progress in
this direction requires to construct new types of quantum field
theories with internal Lie superalgebra symmetries. The precise
model to be considered depends on the particular approach that is
employed. Recent investigations have been based on certain gauge
fixed versions of the Green-Schwarz superstring
\cite{Metsaev:1998it,Pesando:1998fv,Kallosh:1998nx,Pesando:1998wm,%
Rahmfeld:1998zn,Park:1998un}, the pure spinor formalism \cite{Berkovits:2000fe,%
Berkovits:2000yr,Berkovits:2002zv,Berkovits:2004xu} and the
hybrid formalism  \cite{Berkovits:1999im,Berkovits:1999zq}.
\smallskip

Without much further comment on the precise relation with string
theory (see some remarks below, however), we shall turn our
attention to a particular class of quantum theories with internal
supersymmetries, namely to non-linear sigma models on supergroups.
They are characterized by the following simple action
\begin{equation}
  \label{eq:WZW}
   \mc{S}_{f,k}[S]\ = \ -\frac{1}{2\pi f^2}
   \int_{\Sigma} \!d^2z \;\str  \left(S^{-1}\partial S S^{-1}
  \bar{\partial}S \right) -\frac{k}{12\pi} \int_{\Sigma}\; d^{-1}\str
  \left(\left(S^{-1}dS \right)^3 \right)
\end{equation}
with a suitably normalized supertrace $\str$. Here, $S$ is a map
from the world-sheet $\Sigma$ to some supergroup $G$. We have
weighted the standard kinetic term with a coupling constant $f^2$
and also added a topological Wess-Zumino (WZ) term with
coefficient $k$. For sigma models on bosonic groups, quantum
conformal invariance requires $f^{-2} = k$. Once we have adjusted
the coupling constants in this way, we are dealing with a
Wess-Zumino-Novikov-Witten (WZNW) theory which can be solved using
the algebraic techniques of 2-dimensional conformal field theory,
exploiting the infinite dimensional current algebra symmetry of
the WZNW model.
\smallskip

It is one of the intriguing features of certain supersymmetric
target spaces that the requirement of quantum conformal invariance
may not impose any restriction on $f^{-2}$, see e.g.\
\cite{Berkovits:1999im,Bershadsky:1999hk,Kagan:2005wt,Babichenko:2006uc}.
This happens whenever the supergroup $G$ has vanishing dual
Coxeter number. The latter condition is satisfied e.g.\ for the
superconformal groups PSL(N$|$N) that appear in the AdS/CFT
correspondence, but also for OSP(2N+2$|$N) and $D(2,1;\alpha)$. In
these cases, the action \eqref{eq:WZW} gives rise to a continuous
family of conformal quantum field theories. All models share the
same global target space symmetries. On the other hand, the WZ
point with $f^{-2} = k$ is still distinguished by an enhancement
of world-sheet symmetries. For generic values of $f$, one only
expects to find a few chiral higher spin fields in addition to the
Virasoro symmetry that comes with conformal invariance (see
\cite{Bershadsky:1999hk} for details). Whatever the precise chiral
symmetry is, it will almost certainly not suffice for a full
algebraic solution of generic supergroup sigma models. This
insight has lead many scientists working in the field to discard
conformal field theory techniques and to turn to other methods in
integrable systems, such as the Bethe-Ansatz and generalizations
thereof.
\smallskip

Though ultimately, computations in superspace sigma models may
involve a variety of integrable techniques (see e.g.\
\cite{Read:2001pz,Saleur:2001cw,Essler:2005ag,Bena:2003wd,%
MacKay:2004tc,Mann:2004jr,Mann:2005ab,Polyakov:2005ss,%
Beisert:2005tm,Bytsko:2006ut,Teschner:2007ng,Candu:20071}
for an incomplete
collection of recent relevant ideas, a few results and many
further references, in particular to the earlier literature),
it seems to us that the real potential of conformal field theory
methods has not been explored with sufficient care. In fact, we
shall see below that a combination of algebraic techniques with
conformal perturbation theory can provide powerful new results going
far beyond the WZ point. To be more precise, we propose to consider
the sigma models \eqref{eq:WZW} as deformations of a WZNW model,
\begin{equation}
 \mc{S}_{f,k}[S] \ = \  \mc{S}_{k}^{\WZNW}[S] \ - \ \frac{\lambda}{2\pi}
   \int_{\cal H} \!d^2z \;\str  \left(S^{-1}\partial S S^{-1}
  \bar{\partial}S \right) \ = \ \mc{S}_{k}^{\WZNW}[S] +
      \mc{S}_{\lambda}[S]\ \ .
\end{equation}
The deformation parameter $\lambda$ is related to $k$ and $f$
through $\lambda = f^{-2}-k$. For reasons to be explained below,
we shall often refer to this deformation of the WZNW model as a
``RR-deformation''. Note, however, that on the level of sigma
models it simply changes the overall scale factor of the metric
while leaving the magnetic background field invariant. Our
approach is then to study the sigma model through conformal
perturbation theory around the WZ point. In this note we restrict
our attention to the simplest objects, namely to partition
functions, leaving investigations of correlators etc.\ as an
interesting problem for future research.
\smallskip

In order to explain our strategy, let us briefly look at simple
torus compactifications. Suppose we are interested e.g.\ in the
spectrum of strings on a 1-dimensional circle with arbitrary
compactification radius $r$. At generic points in the
1-dimensional moduli space, the chiral symmetry of the model is
generated by the U(1) current $i\partial X$ and its
anti-holomorphic counterpart. With respect to these currents, the
theory is not rational. But there exist some distinguished points
in the moduli space at which the chiral symmetry is enhanced and
the theory becomes rational once the additional chiral fields are
taken into account. In particular, the moduli space contains one
point, known as the self-dual radius $r_0 = r_{\text{SD}}$, where
the symmetry gets enhanced to an $\sll$ current algebra at level
$k = 1$. At this special radius, all spectra can be composed from
a finite number of sectors. With later generalizations in mind, we
consider the partition function on a strip or half-plane with
Neumann boundary conditions which is simply given by the vacuum
character of the $\sll$ current algebra\footnote{At the self-dual
radius there is no fundamental difference between a D-instanton
and an extended brane since they can can be rotated continuously
into the other, see e.g.\ \cite{Recknagel:1998ih}.}
\begin{equation}
Z^{r_0}_{N} (q) \ = \chi^{\text su(2)}_{0,k=1} \ = \
  \vartheta_3(q^2)/\eta(q) \ = \
 \frac{1}{\eta(q)} \ \sum_{n\in \mathbb{Z}} \, q^{n^2} \ \ .
\label{sdPF}
\end{equation}
Other points in the moduli space may be reached through a
deformation with the perturbation ${\cal S}_\gamma =
\frac{\gamma}{2\pi} \int d^2 z \partial X \bar \partial X$. The
perturbation series for the conformal dimensions of boundary
fields can be summed up to all orders in perturbation theory. Our
partition function \eqref{sdPF} gets deformed to
\begin{equation} \label{rPF}
Z^{r}_{N} (q) \ =\
 \frac{1}{\eta(q)} \ \sum_{n\in \mathbb{Z}} \, q^{\frac{n^2}{1-\gamma}}\ \ .
\end{equation}
The result corresponds to the spectrum of a point-like brane on a
circle with radius $r = r_{0} \sqrt{1-\gamma}$. In the
perturbative treatment, the factor $1/(1-\gamma)= 1 +
\gamma/(1-\gamma) $ arises from a geometric series as explained
e.g.\ in \cite{Schomerus:1999ug}. Bulk spectra can also be
computed, either directly or through modular transformation of the
boundary partition function. Let us point out that the
perturbative analysis is insensitive to the fact that the theory
ceases to be rational once we move away from the self-dual radius.
Of course, in this particular case the U(1) current algebra
symmetry is sufficiently large for an algebraic construction of
the theory at generic radii and such a construction is about as
difficult as it is at the self-dual point. Hence, there is no good
motivation to pass through a perturbative construction.
\smallskip

But there exists a better example to illustrate the enormous
potential conformal perturbation theory may possess. It is
provided by the 1-dimensional boundary sine-Gordon theory. In this
model, a periodic potential is switched on along the boundary of a
free field theory. As a consequence, the spectrum of boundary
dimension develops gaps which can grow with the strength $\lambda$
of the perturbation. Eventually, only a point-like spectrum
remains. Given the complexity of the spectrum at intermediate
values of $\lambda$, one might suspect that its precise form is
very difficult to determine. Yet, the boundary partition function
can be calculated rather easily in perturbation theory
\cite{Callan:1994ub,Recknagel:1998ih,Gaberdiel:2001zq}, for any
value of the deformation parameter $\lambda$. In this example, the
boundary potential reduces the chiral symmetry to the Virasoro
algebra. In principle, the latter is still sufficiently large to
allow for a standard CFT construction of the boundary sine-Gordon
theory, but such an analysis is of the same level of difficulty as
the solution of Liouville theory and it has never been carried
out. Hence, the example of boundary sine-Gordon theory supports
our claim that in some situations, conformal perturbation theory
provides an easy route to complicated results that seem (almost)
inaccessible through the usual algebraic methods. A similar
picture will emerge from our study of boundary spectra on
supergroup $\sigma$-models.
\medskip

Even though most of the ideas and technical steps we are about to
explain hold quite generally, we shall carry them out in a
particular example, namely for the supergroup $\PSL$. This allows
our presentation to be very concrete. Furthermore, our results
apply to string theory in $ AdS_3 \times S^3 $ whose solution has
been reduced to the construction of sigma models on the supergroup
$\PSL$ through the hybrid approach developed by Berkovits, Vafa
and Witten \cite{Berkovits:1999im}. In this context, the WZNW
model corresponds to a background with pure NSNS 3-form flux.
Switching on an additional RR field is modelled by the marginal
perturbation with ${\mc S}_\lambda$ which is why we often refer to
this term as RR-deformation. Sigma models on $\PSL$ and closely
related target superspaces have been investigated by several groups
\cite{Berkovits:1999im,Bershadsky:1999hk,Zirnbauer:1999ua,
Guruswamy:1999hi,Bhaseen:1999nm}. For our analysis, the studies
by Bershadsky et al.\ have been particularly useful.
\smallskip

\begin{figure}
  \centerline{
  \raisebox{-25pt}{
  \begin{picture}(80,120)
    \put(0,5){ \scalebox{.8}{\includegraphics[height=7cm]{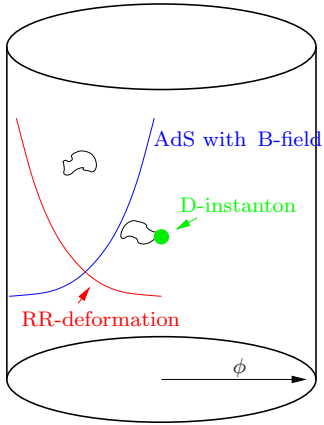}} }
    \setlength{\unitlength}{1pt}
    \put(10,42){\textcolor{red}{\scriptsize RR-deformation}}
    \put(60,110){\textcolor{blue}{\scriptsize AdS with\\\, B-field}}
    \put(70,85){\textcolor{green}{\scriptsize D-instanton}}
    \put(90,24){$\scriptstyle\phi$}
  \end{picture}
  }
  }
  \caption{The influence of NSNS and RR potentials on strings in the bulk and on an instantonic brane.}
\end{figure}

With the example of strings in $AdS_3$ in mind, we may re-evaluate
our optimistic hopes to compute exact spectra through perturbation
theory. Let us think of the target space as a 3-dimensional solid
cylinder. Since $AdS_3$ is curved, the corresponding sigma model
is interacting. At the WZ point, the interaction falls off
exponentially towards the boundary of the cylinder. This has
several effects on the bulk spectrum. In particular, the spectrum
is continuous and there exist so-called long string states that
can stretch along the boundary \cite{Maldacena:2000hw}. The
RR-deformation now adds
another term to the interaction which increases exponentially near
the boundary. Obviously, such a new term must have drastic effects
on the spectrum. Certainly, long string states disappear. In
addition, the spectrum is expected to become discrete since closed
strings are now moving in a box between the two exponential walls.
The dramatic effects of the RR-deformation may raise doubts that
perturbative computations could be successfully performed. And
indeed, it is
most likely true that the bulk spectrum of the theory is not
amenable to a perturbative expansion in $\lambda$. But the situation
changes if we consider the boundary spectrum \cite{Giveon:2001uq,%
Ponsot:2001gt} on a D-instanton instead. Suppose, the instanton
has been placed at the center of the solid cylinder. Open strings
that end on such a D-instanton must be very highly excited in
order to penetrate into the region close to the boundary where the
RR-background flux can be felt. Therefore a D-instanton spectrum
might be accessible through a perturbative computation. Below we
shall see that this intuition is indeed correct. In fact, we are
able to determine the exact spectrum of a D-instanton for any
value of $\lambda$. The same calculation fails at one step when we
try to apply
it to the bulk or to spectra on non-compact branes.%
\smallskip%

Our main new result is a complete computation of the
boundary spectrum for maximally symmetric, point-like branes in
sigma models on the supergroup $\PSL$. The partition function of
such a system was argued in \cite{Gotz:2006qp} to be of the
general form
\begin{eqnarray} \label{Casdec}
& & \hspace*{-2cm} Z_{DI;\lambda}^{\PSL}(z_1,z_2;q) \ = \
\str_{{\cal H}}
   \left(\, q^{L_0-\frac{c}{24}}\, z_1^{K^0_1}\, z_2^{K^0_2}\,
\right)\\[2mm]
  & = &  \sum_{j_1\neq j_2}\, a^\lambda_{j_1j_2}(q) \,
  \chi_{[j_1,j_2]}(z_1,z_2) + \ \sum_{j}\, b^
\lambda_j (q) \,\chi_{{\cal P}[j]}(z_1,z_2)\ \ .  \nonumber
\end{eqnarray}
Here, $K^0_1$ and $K^0_2$ are two Cartan elements in the bosonic
subalgebra $\sll \oplus \sll$ of $\psl$ and we have denoted the
characters of the contributing finite dimensional $\psl$
representations by $\chi$ (see Appendix~A for explicit formulas).
The branching functions $b_j$ and $a_{j_1,j_2}$ at the WZ point
$\lambda =0$ were also determined in \cite{Gotz:2006qp}. Our aim
in this work is to show that the branching functions $b_j$ are
independent of the deformation parameter $\lambda$ while
\begin{equation} \label{main}
a^\lambda_{j_1,j_2}(q) \ = \ q^{-\, C^{j_1,j_2}_2\,
\frac{\lambda}{k(k+\lambda)}} \ a^{\lambda=0}_{j_1,j_2}(q)\ \ \ \
\text{with} \ \ \ \ C^{j_1,j_2}_2 \, = \, j_2(j_2+1) - j_1(j_1+1)\
.
\end{equation} Let us already point out that the dependence of the
conformal weights on the deformation parameter $\lambda$ is very
similar to the one found in free field theory (see eq.\
\eqref{rPF}). We shall see that this is due to some peculiar
features of the Lie superalgebra $\psl$.
\smallskip

Our formulas \eqref{Casdec} and \eqref{main} contain a surprising
wealth of information. Let us unravel some of that through a few
selected cases. Consider, for example, the boundary current
$J^\mu(x)$ where $\mu$ runs through some $14$-dimensional basis of
$\psl$. Under the action of the global $\psl$, the currents
transform in the adjoint representation which is part of the
atypical module ${\cal P}[0]$ (see Appendix~A). Since the branching
functions $b_j$ are independent of $\lambda$, states transforming
in any of the ${\cal P}[j]$ do not receive corrections. Hence,
the currents $J^\mu$ continue to possess dimension $h=1$, as
expected. Things become more interesting once we proceed to
products $J^\mu J^\nu$ of currents. These form a $196$-dimensional
subspace of fields transforming in the $48$-dimensional
representations $[0,1], [1,0]$ and various subspaces of ${\cal
P}[j]$. Hence, under the deformation, the weight of $96$ fields
gets lifted while $100$ fields remain at conformal weight $h=2$.
\smallskip

Formula \eqref{main} passes a few interesting test. To begin with,
we observe that the energy shift is positive for states with
sufficiently large momentum $j_1$ in the radial direction of
$AdS_3$. This is in line with our geometric intuition: Only states
that are highly excited in the radial direction can penetrate to
the region near the boundary of $AdS_3$ where their energy gets
lifted due to the RR perturbation. It is also interesting to
evaluate our formula in the semi-classical regime, i.e.\ for large
values of the level $k$. Inserting the relation $\lambda = f^{-2}
- k$ in \eqref{main} and sending $k$ to infinity, the spectrum of
boundary conformal weights is seen to coincide with the spectrum
of $f^{2} C_2$ up to the usual integer shifts. The eigenvalues of
$f^2 C_2$ may be interpreted as energies for a particle moving on
$\PSL$. Hence, at large level $k$ and modulo integers, the spectrum
of the sigma model on $\PSL$ agrees with the minisuperspace
approximation, as it is supposed to. A much more detailed
investigation of the approach to minisuperspace spectra for
supersphere models is included in a very interesting upcoming
paper by Candu and Saleur \cite{Candu:20072}.\smallskip%

The plan of this work is as follows. In the next section we
collect some background material, partly from our earlier paper
\cite{Gotz:2006qp}. This includes a careful discussion of
maximally symmetric, point-like branes in the WZNW model on
$\PSL$. The ones that are relevant for our analysis are located at
the group unit $e$ of the bosonic base and they extend in all
eight fermionic directions. The associated boundary partition
function is discussed in section~2.2 along with more details on
the Casimir decomposition \eqref{Casdec} at the WZ point.
Section~2.3 contains a construction of the perturbing field in
terms of currents and a new proof of its exact marginality. Most
of our novel results are obtained in section~3 which begins with a
few comments on 2-point functions. Section~3.2 lists several
observations concerning the perturbative series generated by ${\mc
S}_\lambda$. We shall show that the RR-deformation, while being
non-abelian and non-constant on $\PSL$ in general, simplifies
drastically in the evaluation of $\psl$ invariant quantities, such
as conformal weights. In fact, the RR-deformation turns out to be
quasi-abelian, i.e.\ its combinatorics is no more complex than it
is for constant shifts of the closed string background fields in a
flat target space. There remains a mixing problem, however, that
we can only overcome when the general results are applied to
boundary conformal weights of a point-like D-instanton. This is
explained in section~3.3 before we combine all our results into an
exact computation of boundary weights, following closely the steps
of a similar computation in \cite{Schomerus:1999ug}. Our
concluding section includes extensive comments on possible
generalizations, applications and consequences.

\section{Collection of background material}

The purpose of the following section is mainly to provide the
background material that our subsequent perturbative evaluation of
boundary partition functions is based upon. In the first part we
gear up to explain the structure of the boundary partition
function we are about to deform. We start with a few comments on
brane geometries in WZNW models on $\PSL$, extending our previous
analysis of branes in the gl(1$|$1) WZNW model
\cite{Creutzig:2007jy}.\footnote{This first subsection is based on
unpublished notes of TC on branes in supergroup WZNW models.} One
of the instantonic D-branes we find, possesses exactly the
spectrum that was anticipated in \cite{Gotz:2006qp}. The full
field theory partition function and its so-called Casimir
decomposition is reviewed in the second subsection. We then turn
to a more detailed analysis of the perturbing field, mostly
following our previous discussion in \cite{Gotz:2006qp}. On this
occasion, we propose a new argument for the exact marginality of
$\Phi$ which differs a bit in character from the original
derivation \cite{Bershadsky:1999hk,Berkovits:1999im}. Rather than
using the vanishing of the dual Coxeter number of $\psl$, our
reasoning exploits the position of $\Phi$ at the bottom of a
logarithmic multiplet (see also \cite{LeClair:2007aj}). Most of
the results we describe below are not new and the impatient or
experienced reader may skip forward to section~3, at least on
first reading.

\subsection{Branes at the WZ point: gluing conditions and geometry}

As we shall explain in great detail below, the success of our
subsequent exact computation of a boundary partition function for
the sigma model on $\PSL$ hinges on three key properties of the
imposed boundary condition. To begin with, it (i) must preserve
some combination of left and right regular $\psl$ transformations.
At the WZ point, maximally symmetric boundary conditions are
associated with so-called twisted conjugacy classes (see
\cite{Alekseev:1998mc} and \cite{Creutzig:2007jy} for the
supersymmetric case). Explicit formulas for the deformation of the
partition function can only be found if (ii) the corresponding
twisted conjugacy class is point-like localized on the bosonic
base and (iii) it is delocalized in all the fermionic directions.
Later we shall rephrase these conditions as inherent features of
the boundary conformal field theory. Our aim here is to describe a
boundary condition which meets all these requirements and to
determine the relevant boundary partition function at the WZ
point.
\smallskip

In the WZNW model, the global symmetries of the $\PSL$ sigma model
are generated by the zero modes of chiral currents
\begin{equation}
J(z) \ := \ -k \partial S S^{-1} \ \ \ \ , \ \ \ \ \bar J(\bar z)
\ :=\ k S^{-1} \bar \partial S\ \ .
\end{equation}
A boundary WZNW model is scale invariant if the Sugawara stress
tensor obeys $T(z) = \bar T(\bar z)$ all along the boundary $z =
\bar z$. Such a conformal boundary theory preserves a global
$\psl$ symmetry provided that the currents satisfy the following
gluing condition
\begin{equation}
\label{glue} J^\mu (z) \ = \ \Omega \bar J^\mu(\bar z)\ \ \ \mbox{
for } \ z = \bar z \ \ .
\end{equation}
Here, $\Omega$ is a metric preserving automorphism of the Lie
superalgebra. It determines the precise combination $J_0 + \Omega
\bar J_0$ of global $\psl$ charges that remains unbroken by the
boundary condition. In the case of bosonic groups, the geometry
underlying maximally symmetric boundary conditions in WZNW models
was unravelled in \cite{Alekseev:1998mc} (see also
\cite{Gawedzki:1999bq,Felder:1999ka} for various generalizations
and \cite{Schomerus:2002dc} for a review).
There it was shown that a boundary condition in which left and
right moving currents are identified with a trivial gluing
automorphism $\Omega = \id$ correspond to branes whose
world-volume is localized along conjugacy classes. When $\Omega$
is nontrivial, the relevant geometric objects are twisted
conjugacy classes
$$ {\cal C}_u^\Omega \ = \ \{\, h \ \in\  G \, | \, h\ = \
  g u \Omega (g^{-1})\, \} \ \  $$
where $u$ is an element in $G$ and we have lifted the automorphism
$\Omega$ from the Lie algebra to the group. As explained in
\cite{Creutzig:2007jy}, the derivation of \cite{Alekseev:1998mc}
carries over to WZNW models on supergroups (see also
\cite{Creutzig:2007??} for a general analysis).
\smallskip

Having outlined the link between boundary conditions and conjugacy
classes we are now searching for a pair $(u, \Omega)$ such that
${\cal C}^\Omega_u$ meets the requirements (ii) and (iii) we have
listed in the introductory paragraph to this subsection. We shall
not conduct our search systematically. Instead, let us simply
argue that the choice $u = e$ and $\Omega(X) = (-1)^{|X|} X$ does
the job. The corresponding twisted conjugacy class ${\cal
C}_u^\Omega$ is localized at the unit element $e$ of the bosonic
group and it extends in all fermionic directions, i.e.\ along
those tangent vectors $X \in \psl$ which have degree $|X| =1$. It
is easy to see that $\Omega(X) = (-1)^{|X|} X$ is consistent with
the Lie superalgebra structure and the metric. Hence, it extends
to a gluing automorphism on the entire current algebra. Moreover,
parametrizing elements $g$ of the supergroup in the form $g = \exp
(F) \exp(B)$ where $F(B)$ is any linear combination of odd (even)
elements, we find
$$ {\cal C}_u^\Omega \ = \ \{ \, h \ \in \ G \, | \,
   h \ = \ e^F e^B \Omega\left( e^{-B} e^{-F}\right) \ = \
   e^{2F} \, \}\ \ .
$$
Indeed, the bosonic coordinates have dropped out and we remain
with a superconjugacy class of superdimension $0|8$ which extends
merely along the $8$ fermionic directions. We conclude that the
space of functions on the corresponding brane is given by
\begin{equation} \label{gst} f \ = \ f(\eta_a,\bar \eta_b) \ \
\end{equation} where $\eta_1, \dots, \eta_4$ and their bared counterparts
are four fermionic coordinates that parame\-trize the odd generators
$F$. The relevant action of $\psl$ on this $2^8$-dimensional space
is obtained by restricting the $\Omega$-twisted adjoint action of
$\psl$ on the supergroup $\PSL$ to a point in the bosonic
submanifold, or more explicitly,
\def\del{\partial}
\begin{equation}
       \begin{split}
            A^a_{1}  &= \ \bar{\del}^a -
            \frac{1}{2}\epsilon^{abcd}\eta_b(\eta_c\del_d-\eta_d\del_c)-
                     \epsilon^{abcd}\eta_b(\bar{\eta}_c\bar{\del}_d-
                 \bar{\eta}_d\bar{\del}_c) \ , \\[2mm]
            A^{ab}  &= \ -i\eta^a\del^b+i\eta^b\del^a-
                  i\bar{\eta}^a\bar{\del}^b+i\bar{\eta}^b\bar{\del}^a
           \ \ \ \ \  ,\ \ \ \ \
            A^{a}_{2}  \  = \ \del^a \ . \\[2mm]
        \end{split}
\end{equation}
The curious reader can find explicit formulas for the generators
of the left and right regular action $L_X$ and $R_X$ in \cite{Gotz:2006qp}.
When combined as $A_X=L_X + (-1)^{|X|} R_X$, they result in the twisted
adjoint action that is relevant for the minisuperspace description of
the symmetry of our brane.
\smallskip

Under the twisted adjoint action $A_X$, the $2^8$-dimensional
space of ground states \eqref{gst} may be seen to transform
according to the representation
$$ {\cal B}(0,0) \ :=\  {\rm Ind}_{\g^{(0)}}^{\g}
  V_{(0,0)} \ = \ \mc{U}(\g) \otimes_{\g^{(0)}} V_{(0,0)} \
\ \cong \ \P_0 \ \oplus \ [1,0]\ \oplus\ [0,1] \ \ .
$$
Here, $\g^{(0)}$ denotes the bosonic subalgebra of the Lie
superalgebra $\g = \psl$ and we introduced $V_{(0,0)}$ for the
trivial 1-dimensional representation of $\g^{(0)}$. According to
general mathematical results, the module ${\cal B}(0,0)$ is
projective. Hence, it is guaranteed to decompose into a direct sum
of projective modules. The corresponding decomposition is spelled
out on the right hand side. Here, the symbols $[0,1]$ and $[1,0]$
denote 48-dimensional irreducible typical representations (long
multiplets) of $\psl$. These are generated from the two
3-dimensional representations of $\sll \oplus \sll$ by the
application of four fermionic generators. In addition, there
appears the 160-dimensional projective cover ${\cal P}[0]$ of the
trivial representation $[0]$. It is an indecomposable
representation that is built up from irreducible atypicals (short
multiplets) of $\psl$ according to the following diagram
$$ {\cal P}[0]: \ [0] \ \longrightarrow \
  3[1/2] \ \longrightarrow \  2[1] \oplus 6 [0] \ \longrightarrow \
  3[1/2] \ \longrightarrow \
 [0] \ \ . $$
This so-called composition series tells us that ${\cal P}[0]$
contains the trivial representation $[0]$ as a true
subrepresentation. Its representation space is spanned by the
unique invariant element in ${\cal P}[0]$. We call this
subrepresentation $[0]$ the socle of ${\cal P}[0]$. At the other
end of the diagram, i.e.\ in the so-called head of ${\cal P}[0]$,
we find another copy of $[0]$. It is associated with the factor
space of ${\cal P}[0]$ which is obtained if we divide the
projective cover by its maximal non-trivial subrepresentation. A
brief summary of the representation theory of $\psl$ is provided
in appendix A. Many more details can be found in
\cite{Gotz:2005ka,Gotz:2006qp}. We advise readers who are
unfamiliar with indecomposable representations of Lie
superalgebras to consult those references or other mathematical
literature.

\subsection{Boundary partition function and its Casimir
decomposition}

After this brief discussion of brane geometry and the space of
ground states, let us analyze the excited states which arise
through application of current algebra modes. By construction,
these states transform in representations that emerge from a
product of a projective module with some power of the adjoint and
which, by abstract mathematical results, can be decomposed into
projectives. Explicit formulas for the involved characters were
provided in \cite{Gotz:2006qp}. Since we do not need the details
below, we refrain from reproducing these formulas here. In
\cite{Gotz:2006qp} we also explained how sectors erected over
projective modules can be decomposed into representations of the
Lie superalgebra $\psl$. The result can be expressed in the form
\begin{eqnarray}
  \nonumber
  Z_{D0}^{\PSL}(z_1,z_2;q) & = & \chi_{{\cal P}[0]} (z_1,z_2;q) +
  \chi_{[1,0]}(z_1,z_2;q) + \chi_{[0,1]} (z_1,z_2;q) \\[5mm]
  & = &  \sum_{j_1\neq j_2}\, \left(
a^{{\cal P}[0]}_{j_1j_2}(q) + a^{[1,0]}_{j_1j_2}(q)+
a^{[0,1]}_{j_1j_2}(q)\right) \, \chi_{[j_1,j_2]}(z_1,z_2)
 \label{eq:ProjDeco} \\[2mm] & & + \ \sum_{j}\,
\left(b^{{\cal P}[0]}_j (q) +  b^{[1,0]}_j(q)+
b^{[0,1]}_j(q)\right)
     \,\chi_{{\cal P}[j]}(z_1,z_2) \nonumber
\end{eqnarray}
where $\chi_{[j_1,j_2]}$ and $\chi_{{\cal P}[j]}$ are
supercharacters of the Lie superalgebra $\psl$ (see Appendix~A for
explicit formulas). Formula \eqref{eq:ProjDeco} is known as the
Casimir decomposition of the partition function. The various
branching coefficients $a_{i j}$ and $b_j$ count how many times a
projective $\psl$ multiplet appears on a given energy level. These
numbers may be determined with the help of the Racah-Speiser
algorithm. A detailed explanation can be found in
\cite{Gotz:2006qp} along with a few explicit expressions for the
branching of the representation $\aP[0]$. Here it suffices to
recall that the lowest conformal weight $h^{\varpi}_{j_1,j_2}$
among all the multiplets $[j_1,j_2]$ that are generated out of
ground states in the representations $\varpi \cong {\cal P}[0],
[0,1], [1,0]$ satisfies
$$   h^{\varpi}_{j_1,j_2} \ = \ C_2(\varpi) /k + n(j_1,j_2) \ \
\ \mbox{with}  \ \ n(j_1,j_2) \ \in \ \mathbb{N} \ \  $$ where we
denoted the eigenvalue of the quadratic Casimir element in the
representation $\varpi$ by $C_2(\varpi)$.\footnote{The Casimir
element is non-diagonalizable in ${\cal P}[0]$. Its generalized
eigenfunctions possess vanishing eigenvalue.} The same formula
with $j_1 = j_2$ applies to the projective covers ${\cal P}[j]$.
Note that at the WZ point the spectrum has huge degeneracies
because many different representations of $\psl$ can appear on the
same level of the state space. We shall see how the RR-deformation
partially removes this degeneracy.

\subsection{The RR-perturbation and its exact marginalilty}

The most important actress of this work certainly is the
perturbing field $\Phi$ that generates the deformation away from
the WZ point. So, it is important to fully appreciate its
structure and properties. The following discussion is mostly
borrowed from our paper \cite{Gotz:2006qp} which in turn was based
upon \cite{Bershadsky:1999hk,Berkovits:1999im}. The deformation we
are interested in is generated by the field
\begin{equation}
   \Phi(z,\bar z) \ = \ :\str  \left(S^{-1}\partial S S^{-1}
  \bar{\partial}S \right): \ = \ - \frac{1}{k^2}\, :J^\mu(z) \,
  \phi_{\mu\nu}(z,\bar z)\,
  \bar J^\nu(\bar z):
\end{equation}
The second formulation involves the left and right invariant
(anti-)holomorphic currents $J^\mu(z)$ and $\bar J^\nu(\bar z)$
along with some degenerate primary fields $\Phi_{\mu\nu}(z,\bar
z)$ that transform in the (atypical) adjoint representation
$[1/2]$ of $\psl$, i.e.
\begin{eqnarray} \label{JPOPE}
J^\mu(z)\,  \phi_{\nu\rho}(w,\bar w) & = &
\frac{i{{{f^\mu}_\nu}\,^\s}}{z-w}
 \ \phi_{\s\rho}(w,\bar w) + \dots  \ \ \ , \\[2mm]
\bar J^\mu(\bar z) \, \phi_{\nu\rho}(w,\bar w) & = &
\frac{i{f^{\mu\s}}_\rho}
 {\bar{z}-\bar{w}} \ \phi_{\nu\s}(w,\bar w) + \dots\ \ .
\end{eqnarray}
Let us stress that the vertex operators $\phi_{\mu\nu}$ possess
zero conformal weight, as all vertex operators that are associated
with the atypical sector of the theory. According to
\cite{Bershadsky:1999hk}, the field $\Phi$ generates a truly
marginal perturbation $S^\Phi_\lambda$ of the WZNW model. By
construction, the field $\Phi$ has conformal weights $h = \bar h =
1$ but in principle its dimension could change when we perturb the
theory, i.e.\ $\Phi$ could be marginally relevant. This is not the
case. We shall establish true marginality of $\Phi$ through a new
argument, simpler and of a somewhat different character than the
one used in \cite{Bershadsky:1999hk}.
\smallskip

Our key observation is that all $N$-point functions of $\Phi$
vanish identically. Recall from \cite{Gotz:2006qp} that the entire
bulk spectrum of the $\PSL$ WZNW model is organized in projective
modules with respect to global $\PSL$ (left or right)
transformations. Hence, our perturbing field $\Phi$ is part of an
indecomposable $\PSL$ multiplet ${\cal P}[0]$. Since $\Phi$ is
invariant, i.e.\
$$ X \Phi(z,\bar z) \ :=\  [J^X_0,\Phi(z,\bar z)] \ = \ 0 \ \ \ \
\text{ for all } \ \ X \ \in\  \psl\ , $$ it is associated with
the bottom (socle) of the projective cover ${\cal P}[0]$.
Consequently, there must exist another bulk field $\Psi =
\Psi(z,\bar z)$ along with a (fermionic) symmetry generator $Q$
such that
$$\Phi(z,\bar z) \ = \ Q \Psi(z,\bar z) \ = \
  [J^Q_0,\Psi(z,\bar z)]\ \ . $$
The rest of the argument is now rather standard. Let us consider
the $N$-point function of $\Phi$. By the previous comment we can
represent one of the $N$ fields as $\Phi = Q \Psi$ and obtain
\begin{eqnarray}
& & \hspace*{-1cm} \bigl\langle \Phi(z_1,\bar z_1) \prod_{\nu=2}^N
   \Phi(z_\nu,\bar z_\nu) \bigr\rangle \ = \
    \bigl\langle Q \Psi(z_1,\bar z_1) \prod_{\nu=2}^N
  \Phi(z_\nu,\bar z_\nu) \bigr\rangle\  \\[2mm]
  & = & \sum_{\nu=2}^N  \bigl\langle \Psi(z_1,\bar z_1)
  \Phi(z_2,\bar z_2) \dots Q\Phi(z_\nu,\bar z_\nu) \dots
  \Phi(z_n,\bar z_N) \bigr\rangle\ = 0 \ \ .
\end{eqnarray}
We have used the $\psl$ invariance of the expectation value to
re-shuffle $Q$ from $\Psi$ to the other fields. The resulting
$N-1$ terms in the second row all vanish because $Q$ now acts on
one of the invariant field insertions $\Phi$. Since all $N$-point
functions of $\Phi$ vanish, there is no need to ever renormalize
the perturbing field. Hence, its scaling dimension remains
unaltered.

\section{Deformation of the boundary partition function}

With the proper preparation from the previous section we now
come to the central aim of this work: To compute the conformal
weights of boundary fields on our point-like brane as we go beyond
the WZ point. After a few remarks on the general structure of
2-point functions we shall discuss several remarkable features of
the RR-deformation for conformal weights. These lead to drastic
simplifications of the relevant perturbative expansions. In fact,
their combinatorics is no more complex than the combinatorics of
radius deformations in torus compactifications! There remains a
mixing problem, however, that we can only overcome for the
boundary spectra of point-like localized branes. The relevant
argument is presented in the third subsection. Finally, all the
pieces are collected and the conformal weights of boundary fields
are computed explicitly, following a closely related computation
in \cite{Schomerus:1999ug}.

\subsection{The boundary 2-point function}

A boundary partition function stores all information about the
conformal weights of boundary fields. The latter are also encoded
in the boundary 2-point functions which is the place from which we
are going to read them off. In logarithmic conformal field
theories, such as the WZNW model on $\PSL$, the 2-point functions
contains additional data that we are not interested in and, in
fact, cannot compute perturbatively. Since the reader may not be
so familiar with these issues, we shall briefly discuss the
general structure of 2-point functions in RR-deformations of the
WZNW model on $\PSL$.%
\smallskip

Let us recall that our boundary conditions were chosen such that
they preserve a global $\psl$ symmetry. This remains unbroken by
the RR-deformation and hence all quantities in the deformed theory
are organized in $\psl$ multiplets. We shall label the boundary
fields by $\Psi^\pi(x)$ with a superscript $\pi$ that refers to
the $\psl$ representation the field transforms in. As we have
reviewed above, boundary fields on our instantonic brane can only transform in projective
modules $\pi$ of $\psl$. These can be either typical long
multiplets or the projective covers of atypical short multiplets.
In the following discussion we do not have to distinguish between
these two possibilities. The form of the 2-point functions is
strongly constrained by the usual Ward identities expressing
conformal invariance and global $\psl$ symmetry,
\begin{equation}
 \langle \, \Psi^{\pi_1} (x_1) \, \Psi^{\pi_2} (x_2) \, \rangle_\lambda
   \ = \
   \frac{1}{(x_1-x_2)^{\Delta_1(\lambda)+
   \Delta_2(\lambda)}}\ C^{12}(\lambda)\ \ .
\end{equation}
Here, the symbol $C^{12}(\lambda)$ denotes an intertwiner from the
carrier space of the tensor product $\pi_1 \otimes \pi_2$ to the
trivial representation. Let us note in passing that the space of
such intertwiners may be multi-dimensional. The objects $\Delta =
\Delta(\Psi^\pi)$ act on the carrier space of the representation
$\pi$. They describe the action of $L_0$ on the field multiplets
$\Psi^{\pi}$. Therefore, they clearly commute with the action of
$\psl$. We may split $\Delta$ into a term that is proportional to
the identity and a nil-potent contribution, $$ \Delta(\lambda) \ =
\ \ h (\lambda) \cdot {\bf 1}_{\pi} \, + \, \delta(\lambda)
$$ where some finite power of $\delta$ vanishes. If the
nilpotent part $\delta$ is non-zero for one of the fields
$\Psi^{\pi_1}$ or $\Psi^{\pi_2}$ then the 2-point function
contains logarithmic singularities. It is important to stress that
all the quantities we have introduced, namely the constants $h$
and the maps $\delta, C^{12}$ depend on the deformation parameter
$\lambda$. For reasons that will soon become clear, we are not
able to say anything useful about the $\lambda$-dependence of
$\delta$ and $C^{12}$. On the other hand, we shall compute
$h(\lambda)$ exactly, to all orders in perturbation theory. For a
field $\Psi = \Psi^\pi$, the results is
\begin{equation} \label{hdef}
 h_\Psi(\lambda) \ = \ h_\Psi(0) -
   \frac{C^\pi_2}{k} \frac{\lambda}{k+\lambda} \ = \
  h_\Psi(0) -
   C^\pi_2/ k + C^\pi_2 \, f^2  \ \ .
\end{equation}
Here, $C^\pi_2$ is the (generalized) eigenvalue of the quadratic
Casimir in the representation $\pi$, i.e.\ $C^\pi_2 = j_2(j_2+1) -
j_1(j_1+1)$ for $\pi = [j_1,j_2]$ and $C^\pi_2=0$ whenever $\pi$
is one of the projective covers ${\cal P}[j]$. Note that the shift
of the conformal weight only depends on the transformation
behaviour of $\Psi = \Psi^\pi$ under the action of $\psl$. The
simple result \eqref{hdef} is rather remarkable. Let us stress
again that the numbers $h(\lambda)$ provide exactly the
information that is encoded in the boundary partition function. In
particular, the trace over state space is blind to any nilpotent
terms $\delta(\lambda)$ so that our ignorance concerning their
$\lambda$ dependence does not really matter as long as we don't
attempt to go beyond computing partition functions. \smallskip

There is one more comment that might be worth adding. As we have
seen in section~2.3 already, logarithmic conformal field theories
contain many vanishing correlators. In particular, suppose that
$\Psi_1$ and $\Psi_2$ are two fields that are associated with
states in the socle of a projective cover. Then their 2-point
function is bound to vanish by the same arguments we explained in
section~2.3. A related observation was made by Bershadsky et al.\
in \cite{Bershadsky:1999hk}. The authors of that work then went on
to  conclude that the conformal weights of fields in atypical
representations could not be read off from their 2-point
functions. We see now that this conclusion is incorrect. For each
field in an atypical multiplet there exists some field such that
the associated 2-point function is non-zero. If we pick $\Psi_1$
from the socle of a projective cover, for instance, then we can
find an appropriate field $\Psi_2$ in the head of the dual
projective cover.

\subsection{Perturbative expansion for conformal weights}

The perturbative computation of $h_i(\lambda)$ may seem like a
daunting task at first, yet alone because of the very involved
combinatorics of perturbation theory in curved backgrounds. In
this subsection we shall list three observations that will allow
us to drop most of the terms in the expansion for conformal
weights. In fact, the terms that can safely be ignored are
precisely the ones that arise from the curvature of $\PSL$. Such
simplifications, however, only apply to computations of $\psl$
invariant quantities such as conformal weights etc. The reader is
warned never to use the rules we are about to derive for
computations of other structure constants. \smallskip

All observations made in this subsection are based on a simple
mathematical result that was first formulated and exploited in the
work of Bershadsky et.\ al.\ \cite{Bershadsky:1999hk}. Consider
some $\psl$ invariant $A$ and suppose that $A$ may be written as
$A = C_{abc} f^{abc}$ where $f^{abc}$ are the structure constants
of $\psl$ and $C_{abc}$ are some numbers. Then $A$ can be shown to
vanish, i.e.\ $A=0$. Since the supporting argument provided in
\cite{Bershadsky:1999hk} lacks a bit of mathematical precision, we
have included a full proof and further discussion in appendix B of
this paper. Bershadsky and collaborators applied the vanishing of
$A$ to a perturbative construction of the $\psl$ invariant
$\beta$-function. We shall exploit the same result in our
computation of the numbers $h_\Psi$ which are $\psl$ invariants as
well. A similar vanishing criterion is not satisfied for
intertwiners $\Delta$ between two indecomposables or for maps $C$
from the tensor product of indecomposables to the trivial
representation (see also further comments in Appendix~B).
Therefore, we are not able to compute the full 2-point function of
boundary fields, as mentioned before.
\smallskip

Let us now apply this mathematical lemma to our computation of
conformal weights. The perturbative treatment we have in mind
requires to evaluate correlators with insertions of the perturbing
field $\Phi$. Recall that $\Phi$ was composed from the vertex
operators $\phi_{\mu\nu}$ and currents $J^\mu, \bar J^\nu$. An
initial step is to remove all the current insertions through
current algebra Ward identities. In the process, pairs of currents
get contracted using
\begin{equation} \label{eq:ob1}
J^\mu(z) \, J^\nu(w) \ = \ \frac{i{f^{\mu\nu}}_\sigma}{z-w} \,
J^\sigma(w) + \frac{k\kappa^{\mu\nu}}{(z-w)^2} + \dots \ \sim \
\frac{k\kappa^{\mu\nu}}{(z-w)^2}
\end{equation}
The first equality is the usual operator product for $\psl$
currents. Since we are only interested in computing the invariants
$h_\Psi$, we can drop all terms that involve the structure constants
$f$ of the Lie superalgebra $\psl$. This applies to the first term
in the above operator product which distinguishes the non-abelian
currents from the abelian algebra of flat target spaces. Here and
in the following we shall use the symbol $\sim$ to mark equalities
that are true up to terms involving structure constants. In
conclusion, we have seen that, as far as the computation of
conformal dimensions is concerned, we may neglect the non-abelian
nature of the currents $J^\mu$. Obviously, this leads to first
drastic simplifications of the perturbative expansion.
\smallskip

Currents are not only contracted with other currents. They can
also act on the vertex operators $\phi_{\mu\nu}$. The relevant
operator product expansions have already been displayed in eq.\
\eqref{JPOPE} when we first introduced $\phi_{\mu\nu}$. With our
new sensitivity for the appearance of structure constants we
notice immediately that these operator products are proportional
to $f$. Hence, we conclude
\begin{equation} \label{eq:ob2}
J^\mu(z) \ \phi_{\nu\rho}(w,\bar w) \ = \ \frac{i
{{{f^\mu}_\nu}}\,^\sigma}{z-w} \, \phi_{\sigma \rho}(w,\bar w) +
\dots \ \sim \ 0 \ \ .
\end{equation}
Consequently, we can simply ignore all terms in which a current
acts on one of the vertex operators $\phi_{\mu\nu}$. In this
respect, $\phi_{\mu\nu}$ does no longer behave like a vertex
operator, but rather mimics the behavior of a constant background
field. \smallskip

Of course, $\phi_{\mu\nu}$ still is a non-trivial field and it
therefore has possibly singular operator products with other
fields in the theory. Such non-trivial operator products of the
fields $\phi_{\mu\nu}$ could threaten a successful computation of
conformal dimension. Here is where a third observation comes to
our rescue. Note that shifts of the insertion point of the field
$\phi_{\mu\nu}$ are controlled by the following operator version
of the Knizhnik-Zamolodchikov equation
\begin{equation} \label{eq:ob3}
\partial_z \phi_{\mu\nu} (z,\bar z) \ = \ \frac{i}{k} \, {f_{\sigma\mu}}^{\rho}
\  :J^{\sigma}(z) \phi_{\rho \nu} (z,\bar z):   \ \sim \ 0\ .
\end{equation}
This means that in computations of invariants we can treat
$\phi_{\mu\nu}$ as a {\em function} of conformal weight zero. Let
us stress again that the operator products of $\phi_{\mu\nu}$ can
certainly contain singularities. Relation \eqref{eq:ob3} only
asserts that all singular terms may be dropped in computations of
conformal dimensions. \smallskip

The rules \eqref{eq:ob1} to \eqref{eq:ob3} are the main results of
this subsection. They will be employed at the end of this section
when we compute boundary conformal weights. Related observations
for the background field expansion of sigma models on PSL(N$|$N)
were formulated in \cite{Bershadsky:1999hk}. A successful
computation of conformal weights requires one more important
ingredient, though, that is novel to our analysis. This is what we
are going to address next.

\subsection{Perturbation of boundary conformal weights}

Our arguments up to this point have made no use of the fact that
we were setting off to compute conformal dimensions of boundary
fields for a very particular boundary condition. In fact,
everything we have stated applies to whatever conformal dimension
we would like to compute, bulk or boundary. But there remains an
issue that we cannot overcome in such a general context. According
to the results of the previous subsection our vertex operators
$\phi_{\mu\nu}$ behave like a matrix of functions rather than
fields. This simplifies things immensely. But even multiplication
with a set of functions can be a rather involved operation which
we would have to diagonalize explicitly on field space before we
could spell out conformal dimensions. In other words, there still
exists a potentially complicated mixing problem to be solved. Here
is where our special choice of boundary conditions comes in. As we
shall see, it is chosen such that we can effectively replace
$\phi_{\mu\nu}$ by a constant. Thereby, the mixing problem
disappears.
\smallskip

While the reasoning to be detailed below is somewhat technical,
the basic idea is rather simple: Before the bulk field
$\phi_{\mu\nu}$ can act on boundary fields, it must be restricted
to the world-volume of the brane. Since our brane is point-like
localized at the group unit of the bosonic base, the restriction
of $\phi_{\mu\nu}$ contains no further dependence on the bosonic
coordinates and hence should have a rather simple action on
boundary fields.

In order to make this geometric intuition more precise, let us
look at the bulk-boundary operator product expansion of the vertex
operator $\phi_{\mu\nu}(z,\bar z)$. As the world-sheet coordinate
approaches the point $x$ on the boundary of the upper half-plane,
we can re-expand the bulk field in terms of operators $\Psi(x)$ on
the boundary. The leading terms of this
expansion read
\begin{equation} \label{eq:Bbexp}
\phi_{\mu\nu}(z,\bar z) \ = \ \frac{1}{|z-\bar z|^{2/k}} \,
C^{[1,0]}\,  \Psi^{[1,0]}(x) + C^{{\cal P}[0]}\, \Psi^{{\cal
P}[0]}(x) \ + \dots  \ .
\end{equation}
On the boundary, the field with smallest conformal weight is the
multiplet $\Psi^{[1,0]}$ that is associated with the ground states
in the 48-dimensional typical representation $[1,0]$. In addition,
there is one multiplet $\Psi^{{\cal P}[0]}$ of fields with
vanishing conformal weight. All other fields possess positive
scaling dimension and we have not displayed them in the expansion.
The structure constants $C^{[1,0]}$ and $C^{{\cal P}[0]}$ are
largely determined by $\psl$ symmetry. Under the action of the
unbroken global $\psl$, the bulk multiplet $\phi_{\mu\nu}$
transforms in the 2-fold twisted\footnote{All tensor products in
this subsection are constructed with the action $X \rightarrow
X \otimes 1 + (-1)^{|X|} 1 \otimes X$ where the second term is
twisted by the gluing automorphism $\Omega$}  tensor product
$[1/2]\otimes_\Omega [1/2]$ of the adjoint representation. Consequently,
$C^{{\cal P}[0]}$ intertwines between $[1/2] \otimes_\Omega [1/2]$
and the projective cover ${\cal P}[0]$ etc.
\smallskip

Let us recall from the previous subsection that, in all
computations of conformal dimensions, the bulk field
$\phi_{\mu\nu}$ behaves like a set of functions on target space.
Thereby, we are allowed to drop all terms from the bulk boundary
operator product \eqref{eq:Bbexp} which contain a non-trivial
dependence on world-sheet coordinates, i.e.\
\begin{equation} \label{eq:Bbexp1}
\phi_{\mu\nu}(z,\bar z) \ \sim \ C^{{\cal P}[0]}\, \Psi^{{\cal
P}[0]}(x)\ \ .
\end{equation}
Here, $\sim$ has the same meaning as before, warning us that the
relation \eqref{eq:Bbexp1} should only be used in computations of
conformal weights. \smallskip

Further progress now requires to turn attention to the intertwiner
$C^{{\cal P}[0]}$ from the twisted tensor product $[1/2]
\otimes_\Omega [1/2]$ to the projective
cover ${\cal P}[0]$. The precise structure of $[1/2]\otimes[1/2] \cong
[1/2] \otimes _\Omega [1/2]$
has been determined in \cite{Gotz:2005ka}. There, the tensor
product was shown to decompose into four indecomposable
representations. These include the typical multiplets $[1,0]$ and
$[0,1]$ along with the trivial representations $[0]$ and one
atypical indecomposable whose socle consists of a single adjoint
$[1/2]$. The result implies that the space of intertwiners from
$[1/2] \otimes [1/2]$ to the projective cover ${\cal P}[0]$ is
1-dimensional. In fact, the only non-trivial intertwiner $C^{{\cal
P}[0]}$ maps the invariant $[0]$ in $[1/2] \otimes_\Omega [1/2]$ to the
socle of ${\cal P}[0]$. Transferred back to our bulk boundary
operator product \eqref{eq:Bbexp1} we conclude that only the socle
of the boundary multiplet $\Psi^{{\cal }[0]}$ can arise. Since the
corresponding boundary operator is the identity field, we conclude
$$ \phi_{\mu\nu} \ \sim \  c_0\, (-1)^{|\mu|} \kappa_{\mu\nu} \, {\bf 1}
    \ \ . $$
Here, we have used that every intertwiner from $[1/2]\otimes_\Omega[1/2]$
to the trivial representation $[0]$ is related to the metric by
$(-1)^{|\mu|} \kappa_{\mu\nu}$ with $|\mu| = |X_\mu|$ as before.
  Since the field ${\phi^\mu}_\nu$ is a quantum analogue of the
  representation matrix ${R_{\text{ad}}(g)^\mu}_\nu$ and since we are
  evaluating the latter at the unit element, $g=e$, we obviously have
  $c_0=1$. Consequently, in all computations of boundary
conformal weights we are allowed to set $\phi \sim 
(-1)^{|\mu|} \kappa_{\mu\nu}$. Let us stress that our arguments rely
heavily on the fact that we analyze the boundary fields on point-like
branes. In particular, we used that there was no boundary field that
transforms in the atypical $[1/2]$ representation.

\subsection{Computation of boundary conformal weights}

Let us now finally harvest the results of our careful analysis in
the previous two subsections. As we have shown in the second
subsection, the perturbation series for conformal dimensions is
identical to the one that appears in an abelian theory with
constant background fields. Put differently, the currents $J^\mu$
and $\bar J^\nu$ behave like $J^\mu\approx-i \sqrt k\partial
X^\mu$ and $\bar J^\nu \approx i \sqrt k \bar \partial X^\nu$ in a
theory of $14$ free fields $X^\mu$. Moreover, the matrix $\phi_{\mu\nu}$
can be treated as if it was a constant, similar to the parameter $\gamma$
we introduced in our brief discussion of circle compactifications
around eq.\ \eqref{sdPF}. Including our choices of normalization,
the precise relation is read off from
$$ - \frac{\lambda}{2\pi} \Phi(z,\bar z) \ = \ \frac{\lambda}{2\pi k^2}\,
   :J^\mu(z)\, \phi_{\mu\nu}(z,\bar z)
  \, \bar J^\nu(\bar z): \ \sim \ \frac{\lambda}{2\pi k} \,
  \phi_{\mu\nu}(x) \partial X^\mu \bar \partial X^\nu \ \ . $$
Here, we have used a lower case $x$ in the argument of $\phi$ in
order to stress that it behaves like a function on target space.
On the other hand, there is no dependence on the fields $X^\mu$.
For our special choice of $\Omega$, the gluing condition
\eqref{glue} mimics Dirichlet boundary conditions for the bosons and
Neumann boundary conditions for the fermions in free field theory,
$$ \partial X^\mu(z,\bar z) \ = \ -(-1)^{|\mu|}
\bar \partial X^\mu(z,\bar z) \ \ \ \mbox{ for } \ \ \ z \ =\ \bar
z\ \ . $$
Putting things together, our setup is essentially identical to the
starting point of the perturbative analysis in \cite{Schomerus:1999ug},
Hence, we can carry over all results from that paper and conclude
that the change of boundary conformal dimensions can be determined
from an effective perturbing bulk field of the form
\begin{equation} \label{effint}
S_\lambda \  \longrightarrow   \frac{\lambda}{2\pi k }
  \int_{\cal H} dzd\bar z \ \left(\frac {1}{k +  (-1)^F \lambda
   \phi }
  \right)_\mu^{\ \rho} \phi_{\rho \nu} J^\mu(z) \bar J^\nu(\bar z)
\end{equation}
where ${\cal H}$ is the upper half-plane and we are no longer
allowed to contract currents among each other or with the matrix
valued fields $\phi=(\phi_{\mu\nu})$. The matrix $(-1)^F$ is defined
by $(-1)^F_{\mu\nu} = (-1)^{|\mu|}\kappa_{\mu\nu}$. To leading order,
the effective perturbation \eqref{effint} agrees with the original
perturbing term. Higher order contributions are encoded in a
factor $k/(k + \lambda\phi(-1)^F)$ that resembles the familiar
$1/(1-\gamma)$ in the circle compactification (see discussion
after eq.\ \ref{sdPF}). The signs in the denominator take care
of the gluing condition we imposed. There are a few remarks we
would like to add. To begin with, note that there is no need for
any normal ordering in the previous formula, just as in free field
theory with constant background fields. Our effective perturbation
\eqref{effint} has rather limit validity, though. While in
\cite{Schomerus:1999ug} the effective perturbation was used to
compute both the change of conformal weights and of 3-point
couplings, our entire derivation here was restricted to conformal
weights! So, the formula \eqref{effint} for the effective
interaction should never be used in computations of structure
constants. Let us finally point out that for the time being we
only assumed that the left and right moving currents satisfy the
gluing condition \eqref{glue}. Therefore, our result holds for all
branes of this gluing type, including those cases in which the
brane extends along some of the bosonic directions.
\smallskip

In the final step we specialize now to the instantonic brane that
is located at the unit element $e$ of the bosonic base. Using our
results from the previous subsection we may then replace the
functions $\phi_{\mu\nu}$ by constants, i.e.\ we insert $\phi =
(-1)^F {\bf 1}$ into the formula \eqref{effint},
\begin{equation} \label{effint1}
S_\lambda \  \longrightarrow  \ \frac{\lambda}{2 \pi k}
  \int_{\cal H} dzd\bar z \ \left( \frac {1}{k+  \lambda}
\right)  J^\mu(z) (-1)^{|\mu|}\bar J^\mu(\bar z) \ \ .
\end{equation}
The change of the boundary conformal weights is determined by the
logarithmic divergence in the regulated 2-point function which in
turn arises from the simple poles of the operator products between
the effective perturbing field and the boundary fields
$\Psi^{\pi}$. With the usual normalizations, the resulting shift
$\delta_\lambda h$ of conformal weights becomes
$$ \delta_\lambda h(\Psi^\pi) \ = \ - 2\pi \left( \frac{\lambda}{2 \pi k}
\frac {1}{k+  \lambda}\ \pi(J^\mu J^\mu)\right)\ = \ - \frac
{\lambda}{k(k+  \lambda)}\ C_2^\pi \ \ .
$$
Note that the factor $(-1)^{|\mu|}$ in the effective perturbation
is absorbed when we relate the anti-holomorphic current $\bar
J^\mu$ with the boundary value of the holomorphic current $J^\mu$.
As a result, we have established the anticipated formula
\eqref{hdef}.

\section{Conclusions and outlook}

In this note we computed the full spectrum on a point-like brane
in sigma models with target space $\PSL$. The result was obtained
by summing explicitly the perturbation series that is generated by
the RR-deformation ${\mc S}_\lambda$. A non-vanishing topological
WZ term was required in our analysis to guarantee that we could
construct the spectrum directly at one point of the moduli space.
We believe that this is merely a technical condition that can be
overcome, at least in many examples (see next paragraph). A very
decisive element was to focus on invariants of a Lie superalgebra
to which the vanishing lemma (see appendix B) applies. This leaves
ample room for generalization to other supergroup and coset spaces
with psl(N$|$N) or osp(2N+2$|$2N) symmetry. As explained in
section~3.2, the vanishing lemma renders the perturbation series
for conformal dimensions quasi-abelian. On the other hand, the
effective perturbing operator \eqref{effint} requires additional
diagonalization whenever $\phi_{\mu\nu}$ is non-trivial. Here, we
circumvented the issue with our special choice of instantonic
boundary conditions which allowed us to replace $\phi_{\mu\nu}$ by
a constant. Finally, to have sufficient control over the boundary
partition function, a Casimir decomposition of the spectrum had to
be performed. Such a decomposition is not always possible - it
needs the brane to stretch out in all fermionic directions. Since
branes in generic positions are fully delocalized along fermionic
coordinates, no serious limitations should arise for
generalizations to other backgrounds. In the following few
paragraphs we shall go through all our assumptions in more detail,
with an emphasis on general structures rather than the specific
model we dealt with above.
\smallskip

To get our perturbative expansion started, we need the exact form
of the boundary partition function at one point of the moduli
space. In many cases, such an {\em initial condition} may come
from a WZNW model. The solution of WZNW models on type~I
supergroups has been addressed in \cite{Quella:2007hr}, based on
similar studies of several concrete examples \cite{Schomerus:2005bf,%
Gotz:2006qp,Saleur:2006tf}. It may be interesting to stress that
a point with
non-abelian current algebra symmetry may exist in the moduli space
even if no topological term appears in the action of the model
under consideration. The simplest example is once more provided by
circle compactification whose world-sheet symmetry gets enhanced
to an su(2) current algebra at the self-dual radius. Similar
phenomena are very likely to occur for many other principal chiral
models on supergroups or cosets. For example, according to an
intriguing conjecture of Candu and Saleur \cite{Candu:20072},
there exists a particular choice of the coupling at which the
principal chiral model on the supersphere $S^{3|2}$ coincides
with a OSP(4$|$2) WZNW model at level $k=-1/2$. In general, such
special points in moduli space and their exact properties are
difficult to detect. But even if no points with current algebra
symmetries are known to exist, exact spectra may still be
accessible with different techniques, such as the
use of lattice constructions etc. (see e.g.
\cite{Read:2001pz,Candu:20071,Candu:20072}).
\smallskip

Once the WZ point (or any other explicitly solvable point) is
under control, we would like to deform the model. In most cases,
summing up an entire perturbation series is a hopeless enterprise.
Still, we have seen that explicit summation is possible for the
RR-deformation of the $\PSL$ sigma model, at least once we focus
on appropriate quantities such as conformal weights of boundary
fields. Drastic simplifications in the combinatorics of the
perturbative expansion resulted from three observations,
\eqref{eq:ob1} to \eqref{eq:ob3}, in section~3.2. None of them is
specific to a target space with $\psl$ symmetry. In fact, the
underlying technical lemma (reviewed in appendix B) is closely
related to the vanishing dual Coxeter number of $\psl$, a property
$\psl$ shares with three families of Lie superalgebras, namely
psl(N$|$N), osp(2N+2$|$N) and $D(2,1;\alpha)$. These describe the
global symmetries of many interesting superspaces, ranging from
odd dimensional superspheres $S^{2N+1|2N}$ to the coset spaces
that are involved in the AdS/CFT correspondence. We wish to stress
that a vanishing $\beta$ function of the deformation and the
quasi-abelianness of the perturbative expansion for conformal
dimensions appear as two sides of the same coin. Indeed, they can
both be traced back to the vanishing lemma.
\smallskip

Let us also point out once more that, even though the perturbation
series simplifies for all spectra, we were only able to exploit
this fact in the case of point-like branes. It seems to us that
the absence of bosonic zero modes might be an important feature
for the success of the computation, but whether it is decisive
remains an interesting open problem. In particular, our brief
discussion of bulk spectra in $AdS_3$ (see introduction) suggests
that the remaining diagonalization for closed string modes could
be more than a mere technical issue. In case the direct
perturbative computation of bulk spectra
turns out to be impossible, one might still be able to find bulk
conformal dimensions indirectly through modular transformation of
boundary partition functions. Approaching the bulk spectrum
through open closed string duality would certainly require
explicit formulas for the branching functions $a(q),b(q)$, going
somewhat beyond their mere algorithmic construction \cite{Gotz:2006qp}.
Another potential hurdle to
overcome are the modular properties of the branching functions
$a(q),b(q)$ which might be difficult to control. Even if this is
not possible in general, the branching functions might well
combine into simpler objects for specific values of the
deformation parameter $\lambda$. At points with an enhanced
world-sheet symmetry one would expect an infinite number of
branching functions to align such that they build the characters
of a larger chiral algebra. The latter could well possess simpler
modular properties. A systematic detection of points with enhanced
symmetry along the line of deformations and the reconstruction of
the bulk spectrum is a promising path for future research.
\smallskip

Two further comments concern the degeneracies we found in our
D-instanton spectra. According to the results in
\cite{Bershadsky:1999hk}, the chiral symmetry of sigma models on
$\PSL$ is generated by the $\psl$ Casimir fields, and hence is
much smaller than the full Casimir algebra, see \cite{Gotz:2006qp}
for more explanation. Here, we found that the degeneracies of the
boundary spectrum are determined by the Casimir decomposition.
Hence, they are larger than one would have expected based on the
chiral symmetry alone. This is a remarkable result which points
towards the existence of some enhanced (possibly non-local)
symmetry, at least for the boundary spectra we were concerned with
in our work. It would certainly be very rewarding to uncover this
symmetry. A second enhancement of degeneracies is found in the
atypical sector of the model. In fact, the conformal weight of
fields transforming in an atypical representation of $\psl$ do not
receive any corrections. Therefore such fields are guaranteed to
possess an integer conformal weight. Similar phenomena have been
encountered in recent work of Read and Saleur \cite{Read:2007qq}.
Following their analysis we believe that the large degeneracy in
the atypical sector may be explained by the combined action of two
commuting symmetries. One of them is the Lie superalgebra $\psl$
of global transformations. The second should be closely related to
the algebra of Casimir fields or some extension thereof.
\smallskip

Results on non-linear sigma models with super target spaces are
currently not directly applicable to strings in AdS geometries
other than via the hybrid approach for $AdS_3$. Nevertheless we
believe that two rather general lessons can be inferred from our
studies. First of all, conformal field theory techniques, and in
particular conformal perturbation theory, can be rather powerful
even in cases when the chiral symmetry is not sufficient to carry
out a full-fledged algebraic construction of the model.
Furthermore, models with a $\psl$ symmetry can be much better
behaved than one would expect after looking at any of their
subsectors. In fact, supposedly simpler subsectors, such as e.g.\
those based on the bosonic $\sll$, can lead to technical problems
that are much more difficult and never encountered in the full
$\psl$ model. In this sense, subsector theories may turn out to be
inappropriate as toy models for the kind of theories we are
ultimately interested in.
\bigskip \bigskip

\noindent {\bf Acknowledgment:} We wish to thank Constantin Candu,
Gerhard G\"otz, Andreas Ludwig, Sylvain Ribault, Hubert Saleur,
J\"org Teschner and Kay Wiese for
comments or discussions during the various stages of this work.
The research of T.Q.\ is funded by a Marie Curie Intra-European
Fellowship, contract number MEIF-CT-2007-041765. We furthermore
acknowledge partial support from the EU Research Training Network
{\it Superstring theory}, MRTN-CT-2004-512194 and from {\it
ForcesUniverse}, MRTN-CT-2004-005104.%

\newpage

\begin{appendix}
\section{The superalgebra $\mathbf{psl(2|2)}$ and its representations}

The Lie superalgebra $\psl$ possesses six bosonic generators
$K^{ab} = - K^{ba}$ with $a,b= 1,\dots,4$. They form the Lie
algebra $so(4)$ which is isomorphic to $\sll\oplus\sll$. In
addition, there are eight fermionic generators that we denote by
$S^a_\a$. They split into two sets ($\a=1,2$) each of which
transform in the vector representation of $so(4)$ ($a=1,\dots,4$)
which is the $(1/2,1/2)$ of $\sll\oplus\sll$. The relations of
$\psl$ are then given by
\begin{equation}
  \label{eq:CR}
  \begin{split}
[K^{ab},K^{cd}] &\ =\  i \left[ \d^{ac}K^{bd}-\d^{bc}K^{ad}-
                        \d^{ad}K^{bc}+\d^{bd}K^{ac}\right] \\[2mm]
[K^{ab},S^c_\c] &\ =\  i \left[ \d^{ac} S^b_\c - \d^{bc} S^a_\c
   \right] \\[2mm]
[ S^a_\a,S^b_\b ] &\ =\  i \,\e_{\a\b}\, \e^{abcd} K^{cd} \ \ .
  \end{split}
\end{equation}
Here, $\e_{\a\b}$ and $\e^{abcd}$ denote the usual completely
antisymmetric $\e$-symbols with $\epsilon_{12}=1$ and
$\epsilon^{1234}=1$, respectively. An invariant metric is
given by
\begin{align}
  \label{eq:Metric}
  \langle K^{ab},K^{cd}\rangle&\ =\ -\epsilon^{abcd}&
  \langle S_\a^a,S_\b^b\rangle&\ =\ -2 \epsilon_{\a\b}\,\delta^{ab}\ \ .
\end{align}
It is unique up to a scalar factor. The signs have been chosen in
view of the real form $\psu$ which is considered in the main text.
In order to define a root space decomposition of $\psl$ we split
the fermions into two sets of four generators
$$ \g^{(1)}_+ =  {\text{span}}\{S_1^a\}\ \ \ , \ \ \
   \g^{(1)}_- =  {\text{span}}\{S_2^a\}\ \ .
$$
As indicated by the subscripts $\pm$, we shall think of the fermionic
generators $S_1^a$ as annihilation operators and of $S_2^a$ as creation
operators.
\smallskip

Finite dimensional projective representations of $\psl$ fall into
two classes. The first one consists of all the long multiplets.
These are labelled by two spins $j_1,j_2$ with $j_1 \neq j_2$ and
their supercharacters read
\begin{equation}
  \label{eq:FullChar}
  \chi_{[j_1,j_2]}(z_1,z_2)
  \ =\ \tr\Bigl[(-1)^Fz_1^{K_1^0}\,z_2^{K_2^0}\Bigr]
  \ =\ \chi_{j_1}(z_1)\,\chi_{j_2}(z_2)\,\chi_F(z_1,z_2)\ \ .
\end{equation}
where $\chi_{j}(z) = \sum_{l=-j}^j z^l$ are the standard
characters for finite dimensional representations of the Lie
algebra $\sll$ and the fermionic factor $\chi_F$ is given by
\begin{equation}
  \chi_F(z_1,z_2)
  \ =\ 4+z_1^1+z_1^{-1}+z_2+z_2^{-1} -2(z_1^{\frac{1}{2}} +
       z_1^{-\frac{1}{2}})(z_2^{\frac{1}{2}} +
     z_2^{-\frac{1}{2}})\ \ .
\end{equation}
Let us also note in passing that the value $C_2([j_1,j_2])$ of the
quadratic Casimir in typical representations may be expressed as
$$ C_2\bigl([j_1,j_2]\bigr) \ = \ j_2(j_2+1) - j_1 (j_1+ 1 ) \ \ \ .
$$
There exists a second class of projective representations ${\cal
P}[j]$ whose members are labelled by a single spin $j$. They are
built up from short multiplets such that their supercharacter
becomes
\begin{equation}
  \label{eq:ProjBos}
  \chi_{{\cal P}[j]}
  \ =\ \Bigl[2\chi_j(z_1)\chi_j(z_2)-\chi_{j+\frac12}(z_1)
   \chi_{j+\frac12}(z_2) -\chi_{|j-\frac12|}(z_1)
\chi_{|j-\frac12|}(z_2)\Bigr]\,\chi_F(z_1,z_2)\ .
\end{equation}
The quadratic Casimir is non-diagonalizable in the projective
covers, with Jordan cells up to rank five. Generalized eigenvalues
of $C_2$ in ${\cal P}[j]$ are well known to vanish for all spins
$j$. In this sense we shall write $C_2\bigl({\cal P}[j]\bigr) = 0 $.

The characters \eqref{eq:FullChar} and \eqref{eq:ProjBos} are
important ingredients in the Racah-Speiser algorithm that
furnishes the Casimir decomposition for the partition function of
a point-like brane, see \cite{Gotz:2006qp} for details.

\section{Derivation of the main vanishing lemma}

Our evaluation of the perturbative expansion for conformal weights
is based on the fact that a $\psl$-invariant $A$ vanishes whenever
it is of the form $A = C_{abc} f^{abc}$. In order to
make our presentation self-contained the vanishing lemma is
derived below. We use this opportunity to clarify a few
unsatisfactory issues in the original argument
\cite{Bershadsky:1999hk}.%
\smallskip%

For the following discussion it is useful to consider $A,C$ and
$f$ as intertwiners rather than a bunch of numbers. By definition,
an invariant $A$ is an intertwiner from the trivial representation
to itself. Similarly, the structure constants $f^{abc}$ may be
considered as an intertwiner from the 3-fold tensor product of the
adjoint $\ad$ to the trivial representation. The possible form of
$\ad^{\otimes3}$ can be severely constrained using results from
\cite{Gotz:2005ka}. The 2-fold tensor product $[1/2]\otimes[1/2]$
contains three irreducible representations
$\mc{I}=[0]\oplus[1,0]\oplus[0,1]$ as well as a more complicated
indecomposable $\pi_{1/2,1/2}^{\text{indec}}$. The tensor product of
$\mc{I}$ with $\ad$ can easily be evaluated. Furthermore, the typical
contributions to $\pi_{1/2,1/2}^{\text{indec}}\otimes[1/2]$ do not
present any obstacle. This results in the decomposition
\begin{equation}
   \begin{split}
     \ad^{\otimes3}
     &\ = \ [1/2]\oplus2\mc{P}_{1/2}\oplus3\bigl([1,0]\oplus[0,1]\bigr)
            \oplus4\bigl([3/2,1/2]\oplus[1/2,3/2]\bigr)\\[2mm]
     &\qquad\oplus\bigl([2,0]\oplus[0,2]\bigr)
            \oplus\cdots
   \end{split}
   \label{eq:ad3}
\end{equation}
   The remaining terms ``$\cdots$'' are the atypical parts in the tensor
   product $\pi_{1/2,1/2}^{\text{indec}}\otimes[1/2]$. They are built
   by combining the following constituents
\begin{equation}
   \bigl\{2[0]_1,2[0]_3,5[1/2]_1,2[1/2]_3,4[1]_2,[3/2]_1,[3/2]_3\bigr\}
\end{equation}
   into a bunch of indecomposable representations.\footnote{The subscript
     refers to an additional $SL(2,\Complex)$ multiplicity, see
     \cite{Gotz:2005ka}.} The precise form of these indecomposables is
   currently not known to us. Nevertheless one can derive analytically
   that their socles can only contain the representations $[0]_1$ and
   $2[1/2]_1$. Due to the self-duality of $\ad^{\otimes3}$, the same
   statement holds for the heads. One can also check that there is no
   true invariant in $\ad^{\otimes3}$, i.e.\ that the head and
   the socle are formed by two different $[0]$'s. The argument rests
   on an explicit construction of the unique invariant state and the
   subsequent proof that it, in fact, lies in the image of the
   quadratic Casimir operator. Hence the unique invariant state
   has to be  the socle of a larger indecomposable multiplet.

   Given any
representation of $\psl$, the number of independent interwiners to
the trivial representation may be obtained by counting the number
of times $[0]$ appears as the head of an indecomposable
sub-representation. In the case of $\ad^{\otimes3}$, there is
only one such occurrence of $[0]$, as we have just
argued. Hence, the intertwiner to the
trivial representation is unique up to normalization. This map is
what we denote by $f$. Bershadsky et al. now continued to argue
that the constants $C_{abc}$ that are contracted with $f^{abc}$ to
form the invariant $A$ must be proportional to $f_{abc}$ (indices
lowered with the metric) because of the uniqueness of $f$. $A$
then vanishes because of the numerical identity $f_{abc} f^{abc}
=0$. We arrive at the same conclusion if we employ that $f$ and
$C$ combine into an invariant $A$ provided that $C$ is a
co-invariant, i.e.\ an intertwiner from the trivial representation
to the 3-fold tensor product of the adjoint. Such co-invariants
are in one to one correspondence with representations $[0]$ in the
socle of $\ad^{\otimes3}$. A glance back onto our argument above
shows that there is a single such representation and hence $C$ is
unique. The reason for the vanishing of any invariant $A = C \circ
f$ is that the image $\text{Im}\,C$ of $C$, given by the socle of
$\ad^{\otimes3}$, is in the kernel of $f$, i.e.\ $\text{Im}\,C$ has no
component in the head of $\ad^{\otimes3}$. The outcome of this
analysis, namely the vanishing of an invariant $A = C \circ f$, is
the crucial ingredient in our observations \eqref{eq:ob1} to
\eqref{eq:ob3}.

\end{appendix}
\bibliographystyle{JHEP-TQ}
\def\cprime{$'$} \def\cprime{$'$}
\providecommand{\href}[2]{#2}\begingroup\raggedright\endgroup

\end{document}